\documentclass[preprint]{revtex4-1}
\usepackage{graphicx}% Include figure files
\usepackage{dcolumn}% Align table columns on decimal point
\usepackage{bm}% bold math
\usepackage[utf8]{inputenc}
\usepackage[T1]{fontenc}
\usepackage{mathptmx}
\usepackage{etoolbox}
\usepackage{xcolor}
\makeatletter
\def\@email#1#2{%
 \endgroup
 \patchcmd{\titleblock@produce}
  {\frontmatter@RRAPformat}
  {\frontmatter@RRAPformat{\produce@RRAP{*#1\href{mailto:#2}{#2}}}\frontmatter@RRAPformat}
  {}{}
}%
\makeatother
\begin{document}

%\preprint{AIP/123-QED}

\title[Wavy-wall-based flow control for airfoil at high Re]{Wavy-wall-based flow control for the suction side geometry of NACA4412 at \boldsymbol{$Re_\tau \approx 3000 $}}
% Force line breaks with \\
\author{A. Dróżdż}
 %\altaffiliation{Department of Thermal Machinery, Czestochowa University of Technology. Armii Krajowej 21, 42-200 Czestochowa, Poland}\\%Lines break automatically or can be forced with \\
\author{M. Romańczyk}%
\author{W. Elsner}
 \email{artur.drozdz@pcz.pl}
\affiliation{ 
Department of Thermal Machinery, Czestochowa University of Technology. Armii Krajowej 21, 42-200 Czestochowa, Poland%\\This line break forced with \textbackslash\textbackslash
}%

\date{\today}% It is always \today, today,

\begin{abstract}
The paper presents a high Reynolds number experimental study of turbulent boundary layer separation control on a convex plate using the wavy-wall method, which was initially proposed for a flat plate by Dróżdż \emph{et al.} 2021 (Exp Therm Fluid Sci 2021;121:110291). The application of this method increases the friction coefficient by up to 42.3\%, resulting in a substantial delay in turbulent separation from the convex wall, while maintaining total momentum, quantified by changes in momentum-loss thickness. Other parameters indicating the high efficiency of the method are the invariant value of the friction Reynolds number along the flow and the thinner boundary layer. The above indicators demonstrate promising aerodynamic improvements in airfoils, similar to those achieved when active suction is applied to the suction side. The new insight into the physical mechanism of the wavy wall suggests that small-scale turbulent activity is the primary determinant of its effectiveness in enhancing small-scale streamwise convection and the sweeping motion, resulting in superior momentum transport. It has been demonstrated that if the wavy wall, due to poorly selected geometry, induces large-scale motions, such as separation in the trough, it counteracts the mechanism induced. Then this geometry has a detrimental effect on the method's peak efficiency, reducing the maximum achievable gains relative to the optimal configuration, while still outperforming the unmodified smooth baseline.
\end{abstract}

\maketitle

\section{Introduction} %%%%%%%%%%%%%%%%%%

Most passive flow control strategies for airfoils are based on reducing skin friction by attenuating large-scale streaks \cite{RICCO2021100713}. To mitigate skin friction reduction, which acts as a laminar-turbulent transition postponer, active methods are employed to counteract its detrimental effect on lift due to earlier separation. However, the use of active methods is limited in wind turbine blades due to a combination of practical, aerodynamic, and economic constraints. The trade-off solution for that problem is not to minimise skin friction drag but to minimise total entropy generation in the boundary layer and wake behind the airfoil by reducing separation occurring for non-optimal inflow conditions. 

Modern multi-megawatt wind turbines operate in high-Reynolds-number aerodynamic regimes ($Re_c \sim 10^{7}$), where maintaining attached flow over the suction side of rotor blades remains a central challenge in maximising aerodynamic efficiency and annual energy yield. Under adverse pressure gradients (APG) characteristic of high angles of attack and off-design operating conditions, the turbulent boundary layer decelerates, thickens, and ultimately detaches from the convex surface. To mitigate these performance penalties, flow control techniques are categorized into active and passive approaches. A commonly applied passive method to prevent flow separation in a wind turbine is the use of vortex generators. The key is that pressure drag dominates once the separation begins. Preventing separation yields a nonlinear reduction in total drag. However, in large-scale wind turbine blades, vortex generators provide only minimal benefits - about a 1\% power increase on 5MW wind turbine blades \cite{Lee2019}. Alternative passive flow control methods such as dimples \citep{tay2015mechanics, gattere2022dimples, aoki2012mechanism, bearman1976golf, tay2011determining}, grooves \citep{Seong2016}, slotted airfoils \citep{belamadi2016aerodynamic, coder2020design, TANURUN2026111604}, or microcylinders \citep{WANG2018101, mostafa2022quantitative, wang2023wake} have been shown to reduce drag by speeding up the transition; however, their effectiveness diminishes in turbulent flows with high Reynolds numbers as shown by McMasters and Henderson \cite{mcmasters1979low}. For multi-megawatt wind turbines, even a 1-2\% reduction in total drag or a 3-5\% increase in lift‑to‑drag ratio results in a significant increase in annual energy production. Active systems—such as localized synthetic jets, boundary layer suction, or unsteady blowing—can adapt dynamically to fluctuating inflow conditions; however, they require parasitic power supplies, complex mechanical integration, and ongoing maintenance, making them difficult to implement on utility-scale turbine blades.

An alternative passive strategy that has attracted increasing attention is the implementation of a solid streamwise wavy wall (WW)—a continuous surface undulation with a viscous-scaled amplitude of $A^+\approx 170$, and a period $ \lambda$ adjusted so that flow troughs are held on the verge of separation (effective slope around 0.15)-designed to modify near-wall turbulent transport. Fundamental investigations into turbulent boundary layer transport over corrugated surfaces originate from the work of Zilker et al. \cite{Zilker1977}, who showed that the flows over $A^+ \approx 27$ cannot be predicted by linear theory. The appropriately selected corrugation geometry at the friction Reynolds number $Re_{\tau}\approx4000$ produces an increase in $C_f$ downstream of the WW of more than 35\% \cite{Drozdz2025}, and can be a remedy for large rotors, where a large portion of the boundary layer is turbulent. This is the highest gain reported by wall modification to date at high Reynolds number, highlighting the practical potential of this method to delay flow separation. To be effective, the WW design must be carefully tailored to local flow conditions. The undulations should be placed where the Rotta-Clauser pressure gradient parameter $\beta=\delta^*\tau_w^{-1}dP_e$, where $\delta^*$ is displacement thickness, $\tau_w$ is wall shear stress and $dPe$ is streamwise pressure gradient, remains below 10. The first, very small, maximum in $C_f$ at $A/ \lambda  \approx  0.03$ over a solid wavy channel was observed in numerical simulations by Yoon \emph{et al.} \cite{Yoon2009} of turbulent flow at a bulk Reynolds number of 6760. Segunda \emph{et al.} \cite{segunda2018experimental} observed that the overall normal Reynolds stress profile increases with the number of wave periods. The most substantial difference between the profiles was observed between the first and second waves, and no further changes in the profiles were observed for $A/\lambda = 0.1$ when the number of waves exceeded 8. Similar observations at $A/\lambda =0.05$ were reported by Hamed \emph{et al.} \cite{hamed2015turbulent}, although a substantial increase in overall Reynolds stress profiles was achieved only over the first three wavelengths. Under these conditions, a relatively strong shear layer is formed upstream of the wave peak, producing a small-scale spanwise vortex, provided that mean flow separation does not occur downstream of the wave peak \cite{Kaminski2024a}. This particular wavy wall increases momentum transport towards the wall due to increased small-scale sweeping motions produced by a propagating faster small-scale spanwise vortex over the downhill slope where an APG is locally strengthened. As it was shown in Dróżdż \emph{et al.} \cite{drozdz2023convection} the faster propagating vortical small-scale structures widely occupy the flow at strong APG conditions and originate from the mechanism of amplitude modulation of small-scale by the large-scales in high-Reynolds-number wall-bounded flows \cite{mathis2009comparison}. According to this mechanism, small-scale turbulence is densely distributed in the high-speed zones near the wall and plays a major role in turbulent transport. The convection velocity is strengthened in the local APG region on the downhill slope only when the flow is still attached there. The increase in momentum as the flow moves towards the wall causes an increase in the flow’s energy there and, as a result, may delay the downstream separation. In the other case, when separation occurs, the WW generates a large pressure drag; therefore, the mechanism underlying the increase in sweep events is too weak to counteract the pressure-drag-induced skin-friction loss. Recently, Opperman \emph{et al.} \cite{Opperman2026}, emphasise that there are no systematic studies of a flow structures over wavy walls in a high Reynolds number turbulent boundary layer that is at the verge of separation in troughs. 
A turbulent boundary layer with high Reynolds number is the primary determinant of the phenomenon of postponement of flow separation by wavy walls, as the approach loses efficacy below a certain threshold Reynolds number, as was shown in combined numerical LES and wind-tunnel studies by Elsner \emph{et al.} \cite{elsner2022experimental} at $Re_{\tau}\approx1400$. For that threshold Reynolds number, the two opposing mechanisms that occur, namely the pressure-drag-induced friction coefficient loss from one side and the sweeping motion friction coefficient gain from the other side, are at a similar level. In numerical calculations using the LES method at a higher Reynolds number of about $Re_{\tau}\approx2500$ Kamiński \emph{et al.} \cite{Kaminski2024a,Kaminski2026} confirmed that such a surface modification can enhance the wall-normal velocity gradient and wall shear stress downstream, although to a lesser extent than in the experiment.

The recent work of Dróżdż \emph{et al.} \cite{Drozdz2025} aimed to test the single geometry of the WW selected in Ref. \cite{drozdz2021effective} on a range of Reynolds numbers ($Re_{\tau}\approx2600$, $4000$, and $4500$) and under varying pressure gradient conditions corresponding to large-scale pitch-regulated wind turbine blades. 
For such turbines, the angle of attack varies from $3^\circ$ to $8^\circ$ \cite{Sayed2012467} for different wind speeds, and even more due to the rotational position or torsional deflection of the blade. 
The results showed that when the Reynolds number varies, as when the wind speeds change from 5 to $40 ms^{-1}$ on a pitch-regulated wind turbine, the lowest increase in $C_f$ was 30\% for the lowest Reynolds number, which is just 5\% less gain than for the optimal design for  $Re_{\tau}\approx4000$ \cite{drozdz2021effective}. Furthermore, when the pressure gradient was varied at a constant Reynolds number, mimicking unstable flow conditions arising from changes in the rotational position of the blade or torsional deflection, the efficiency of the wavy wall was maintained, although at a lower $C_f$ level, about 27\% on average. The tested pressure ranges and the variation in Reynolds number also showed that too-low amplitudes limit the benefits. However, excessively high amplitude leads to significant separation on the downhill slope, thereby weakening the transport of momentum to the wall. Nevertheless, as demonstrated in Dróżdż \emph{et al.} \cite{Drozdz2025}, an overall effectiveness of approximately 23\% for the least beneficial conditions (angle of attack and Re number), was achieved, which can be considered high. 

Recently, large-eddy simulations at $Re_{\tau}\approx2500$ by Kamiński \emph{et al.} \cite{KAMINSKI2025135758} were published as an extended work of Kamiński \emph{et al.} \cite{Kaminski2024a}. They showed that the additional benefit can be achieved with an upstream-tilted wavy wall \citep{KAMINSKI2025135758}. They report an increase in wall shear stress with respect to a sinusoidal type of wavy wall. This increase, however, is lower than in the experimental studies on sinusoidal WW described in \cite{drozdz2021effective}, reaching $\sim 40\%$ of the value observed there; this may be due to the lower assumed Reynolds number. On the other hand, Kamiński \emph{et al.} \cite{KAMINSKI2025135758} presented their observations on the physical interpretation of introducing a wavy wall on the curved surface of the NACA 4412 wing section. They found that a sinusoidal wavy wall used at $Re_\tau \approx 2500$ causes increased energy across all scales in the outer region. While, near the wall, only large-scale activity was enhanced, despite the small-scale contribution observed by Dróżdż \emph{et al.} \cite{drozdz2021effective}. The results of this study \cite{KAMINSKI2025135758} at $Re_{\tau}\approx2600$ showsthat small-scale energy increases when an upstream tilted wavy wall is applied to a curved surface, and this condition primarily influences the effectiveness of the wavy wall.

The paper presents, for the first time, an experimental performance analysis of a solid sinusoidal wavy wall on a convex surface that resembles a wind turbine blade at a high Reynolds number of approximately $Re_{\tau} \approx 3100$, corresponding to a wind speed $10 ms^{-1}$ that occurs on a multi-megawatt wind turbine blade that spins at a constant speed regulated by the blade pitch. The changes in velocity statistics and boundary-layer parameters produced by the wall modification were compared with those obtained using active techniques reported in the literature \citep{Atzori2020}. New insights into the wavy-wall-induced mechanism are provided based on newly obtained data and already published results. This work aims to:
\begin{itemize}
  \item Quantify the local and integrated enhancements in skin-friction coefficient ($C_f$) and the resulting tangential-to-the-surface shift in the boundary layer separation point ($\Delta x_{t,s}/c$).
 \item Elucidate the physical mechanism by which near-wall small-scale sweeping motions maintain momentum transport to the wall while suppressing wall-normal convection into the wake.
 \item Establish a physical termination criterion for wavy surface modifications based on the scale interaction between inner-layer small-scale turbulence and outer-layer large-scale structures.
\end{itemize}
\section{Experimental setup} %%%%%%%%%%%%%%%%%%

\begin{figure*}
	\begin{center}
	\includegraphics*[width=0.618\linewidth]{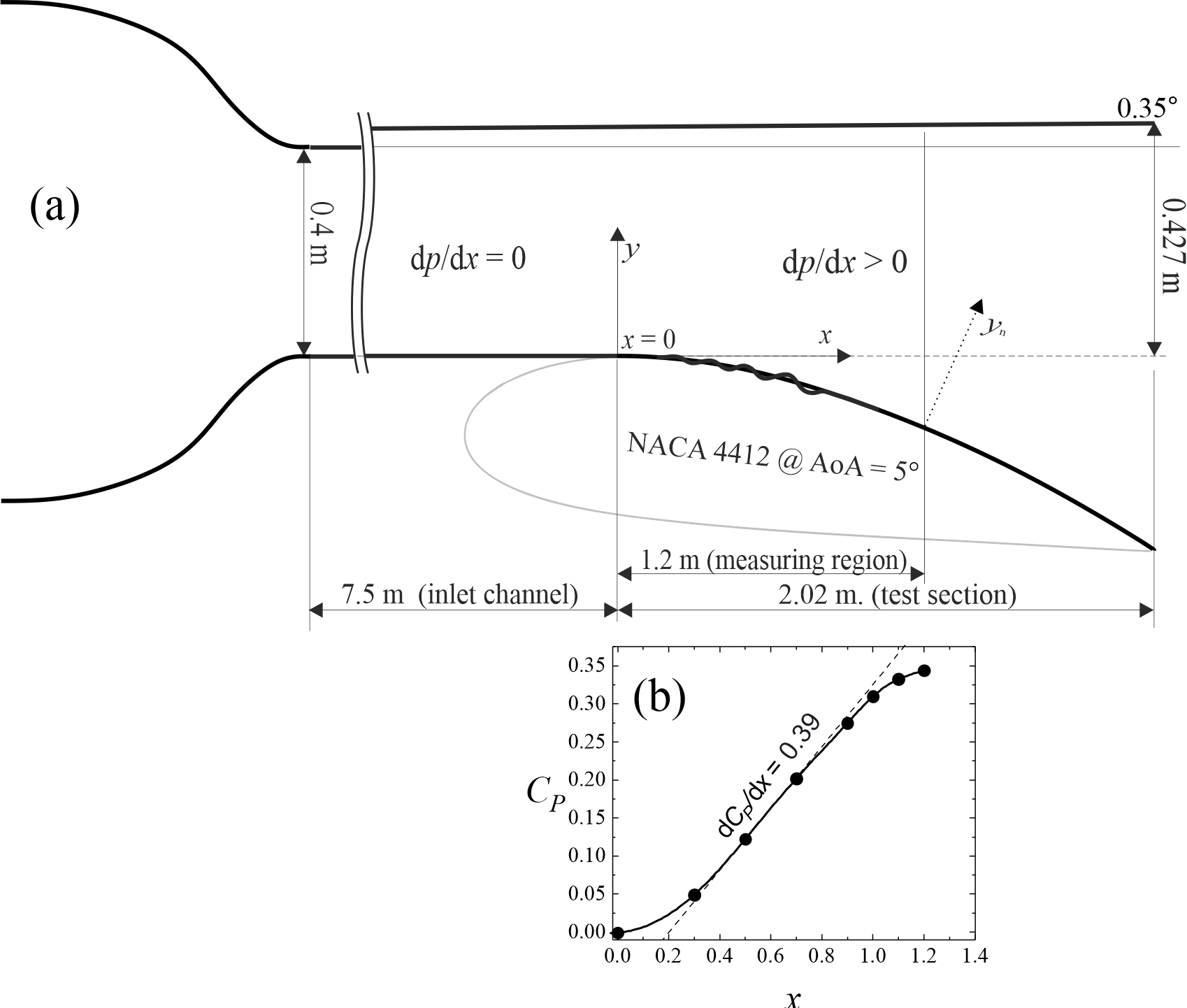}
      
	\caption{\label{fig_stand} Schematic of the test section (a) and $C_p$ distributions (b)}
	\end{center}

\end{figure*}

Experimental investigations were performed using a specialized modular wind tunnel facility at the Czestochowa University of Technology constructed specifically to replicate atmospheric boundary layer conditions over wind turbine blade profiles, where the canonical turbulent boundary layer can develop untripped along a $7500 mm$ flat plate (see Figure~\ref{fig_stand}). The Reynolds number based on developing distance is $Re_x = 10^7$ characteristic for a flow over a 2.0 m chord wind turbine blade under a velocity inlet $80 ms^{-1}$. Such a condition occurs for an average wind speed of about $10 ms^{-1}$. The honeycomb-equipped settling chamber, five turbulence grids, and the contraction channel of contraction ratio of 4.5:1 ensure that the turbulence intensity in the inlet plane was not higher than 0.45\%. To mitigate three-dimensional effects, i.e., secondary flows in the channel, the channel cross-section at the inlet was designed with the following $WxH$ dimensions: $800 \times 416$ mm.
Downstream of the inlet plane, a curved surface section corresponding to the NACA 4412 profile's suction side, positioned at an angle of attack (AoA) of $5^\circ$, was installed (see Figure~\ref{fig_stand}). Zero-pressure-gradient (ZPG) conditions within the initial entry section were maintained by angling the top tunnel wall outward by $0.35 ^\circ$ (see Figure \ref{fig_stand}~a).
The two-dimensional WW section, in which the waviness was extruded along the spanwise direction (800 mm), has a total length of $L=540$ mm ($4$ streamwise periods with $\lambda=135$ mm). The wavy wall was manufactured from a high-density polyurethane plate to ensure a smooth hydraulic surface after mechanical processing. After machining, the plate was bent to match the convex surface.

The amplitude of WW $A^+=A(x)u_\tau\nu^{-1}$, where $u_\tau$ is the friction velocity and $\nu$ is the kinematic viscosity, was tuned to flow conditions occurring in the numerical simulation using LES at $Re_\tau \approx 2600$ (see Kamiński \emph{et al.} \cite{Kaminski2025}). In the experimental setup, it was found that the $A^+$ increases downstream of the flow from $150$ for the first wave up to $170$ for the last  wave. The distributions of the pressure coefficient $C_P=1-U_e^2U_{e,in}^{-2}$ are shown in Figure \ref{fig_stand}b. It should be emphasised that the waviness, due to its low amplitude in relation to the height of the channel, does not affect the pressure distribution in the freestream. The experiment was carried out for the value of the inlet velocity $U_{e, in} \approx 14.5 ms^{-1}$, which corresponds to $Re_ {\tau}\approx 3100$.

The StreamlinePro DANTEC hot wire anemometer (HWA) was used at an overheating ratio of 1.8. Two distinct single-wire probe configurations were deployed: For baseline measurements, the standard gold-plated 55P05 was used, with a wire sensing length of $l=1.2 mm$ and diameter of 5 $\mu m$. In that case, not all profiles are free from spatial averaging (see Table \ref{tab:Re} and Appendix). In the case of a wavy wall, an increase in friction velocity occurs; therefore, a modified 55P31 probe was used, with a wire sensing length of $l=0.41 mm$ and a diameter of 3 $\mu m$. Such wire dimensions ensured maintaining the nondimensional wire length $l^+ = lu_{\tau}\nu^{-1} < 9$.  Data acquisition was performed with sampling frequencies from 25 kHz for profiles taken above $x = 0.9$ m and $50 kHz$ for profiles on the unmodified wing surface up to this location. The temperature variation during the measurement of each velocity profile was not higher than ±0.2 K. The flow temperature was used to correct the voltage signal from the hot-wire.

The friction velocity values $u_{\tau}=\sqrt{\tau_w/\rho}$, where $\tau_w = \mu dU/dy$ and $\mu$ is the dynamic viscosity and $dU/dy$ is the wall-normal streamwise velocity gradient at the wall, were estimated using the approach proposed by Niegodajew \emph{et al.} \cite{NIEGODAJEW2019108456}. The accuracy of this approach was independently verified against oil-film interferometry, yielding maximum friction velocity uncertainties of $2.5\%$ for profiles characterised by shape factor $H < 2.0$ and up to $5.0\%$ for $H > 2.0$. The boundary layer thickness $\delta$ was estimated using a method proposed by Vinuesa \emph{et al.} \cite{Vinuesa2016b}, which is based on the diagnostic plot corrected for APG flows introduced by Dróżdż et al. \cite{Drozdz2014a}. The integral measures ($\delta^*$ and $\theta$) were evaluated by numerically integrating the mean streamwise velocity profiles. To eliminate near-wall thermal conduction artifacts caused by heat transfer from the sensor to the solid surface, velocity measurements below $1.0$ m/s were discarded. The velocity profile in the immediate vicinity of the wall was subsequently reconstructed by applying the linear approximation down to the no-slip condition ($U(0) = 0$). Finally, a trapezoidal numerical integration across the boundary layer thickness ($\delta$) was used. The uncertainties in the measurement/estimation of $U$ and $H$ were below the level of $0.01U_{e,in}$ and $1.5\%$, respectively. Precision positioning of the hot-wire in the wall-normal direction was achieved via optical reflection monitoring using a high-magnification macro-lens camera setup, establishing a wall-distance positional uncertainty $\Delta y_m = 0.06$ mm.

\section{Inlet conditions} %%%%%%%%%%%%%%%%%%

\begin{table}
    \centering
        \caption{Inlet TBL parameters}
    \label{tab:inlet}
    \begin{tabular}{ccccccccc}
        \hline
         & $z$ & $U_{e, in}$ &$\delta_{in}$ & $C_{f,in}$ & $\delta^*_{in}$ & $\theta_{in}$ & $H_{in}$ & $u_{\tau, in}$ \\
       
          & $(m)$& $(ms^{-1})$ & $(mm)$ & $(-)$ & $(mm)$ & $(mm)$ & $(-)$ & $(ms^{-1})$ \\
        \hline
        \textcolor{red}{\scalebox{0.7}{$\bigcirc$}} & -0.2 & 14.58 & 98.4  & 0.00255 & 15.21 & 11.41 & 1.333 & 0.520 \\
        \textcolor{magenta}{\scalebox{0.7}{$\bigcirc$}} & -0.1 & 14.32 &  93.5  & 0.00259 & 14.33 & 10.74 & 1.334 & 0.515 \\
        \scalebox{0.7}{$\bigcirc$} & 0 & 14.33 & 90.5 & 0.00256 & 14.45 & 10.74 & 1.345 & 0.513 \\
        \textcolor{cyan}{\scalebox{0.7}{$\bigcirc$}} & 0.1 & 14.34 & 93.5 & 0.00255 & 14.75 & 11.00 & 1.341 & 0.512 \\       
        \textcolor{blue}{\scalebox{0.7}{$\bigcirc$}} &0.2 & 14.48 & 91.9 & 0.00259 & 14.24 & 10.67 & 1.334 & 0.522 \\
        \hline
    \end{tabular}
\end{table}

\begin{figure*}
	\begin{center}
	\includegraphics*[width=0.9\linewidth]{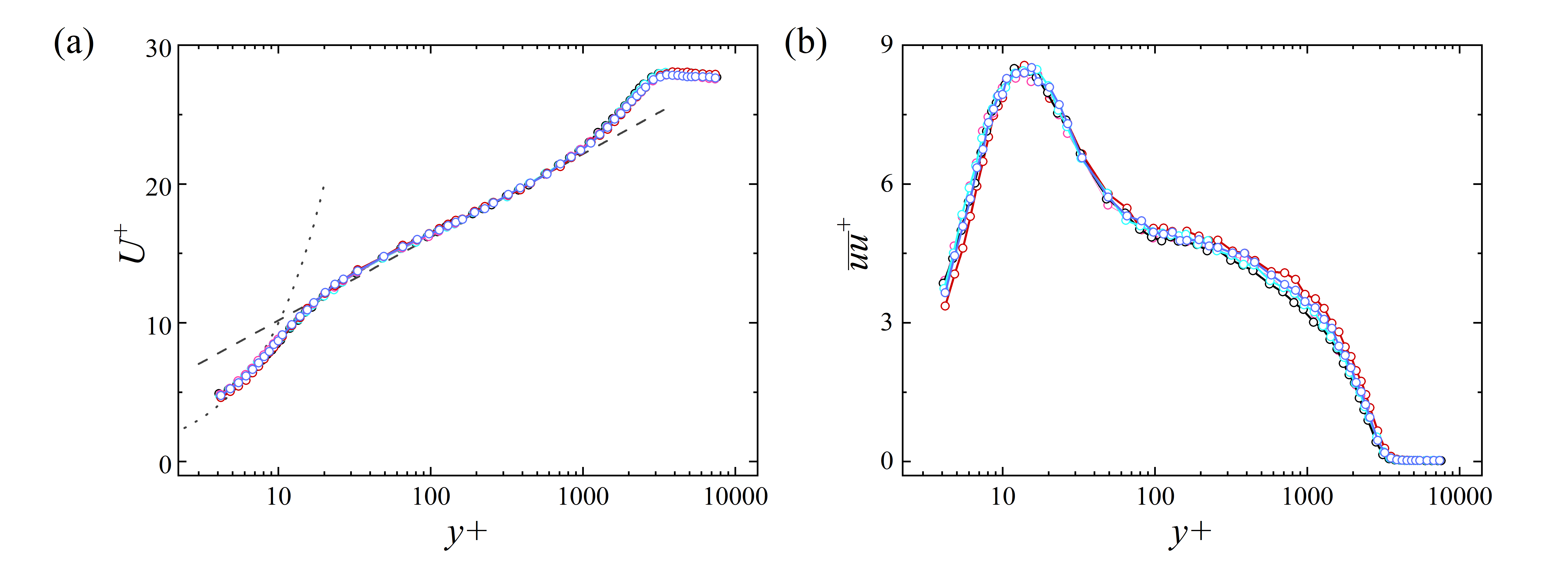}
	\caption{\label{fig_inlet} Mean velocity (a) and turbulence intensity (b) profiles presented in viscous scale at $x = 0$. The dashed line represents the logarithmic law (with $\kappa = 0.381$ and $B = 4.17$) while the dotted line represents $y^+ = U^+$.}
	\end{center}
\end{figure*}
Table \ref{tab:inlet} contains inlet parameters of  five velocity profiles distributed spanwise. For each case, a \scalebox{0.7}{$\bigcirc$} symbol colour has been assigned. It is important to note that the flow maintains well-behaved ZPG TBL flow conditions introduced by Sanmiguel Vila \emph{et al.} \cite{Sanmiguel2017}, as the skin friction coefficient $C_f$ and the shape factor $H $ for analysed $Re_{\theta, in} \approx 9700$ in the inner 50\% of channel span do not exceed ±2\% and ±1 of the Coles–Fernholz formula and the relation proposed by Monkewitz \emph{et al.} \cite{Monkewitz2007}, respectively (see Figure (\ref{fig_inlet}~a)).

Figure \ref{fig_inlet} shows the inlet profiles ($x = 0.0$ mm) of the mean velocity $U^+=Uu_{\tau}^{-1}$ (\ref{fig_inlet}~a) and the streamwise Reynolds stresses $\overline{uu}^+$ in the viscous units (\ref{fig_inlet}~b), measured for five spanwise locations $z = -200$, $-100$, $0.0$, $100$ and $200$ mm. Figure \ref{fig_inlet}~a) shows a canonical turbulent boundary layer at the inlet, while Figure \ref{fig_inlet}~b) shows that $\overline{uu}^+$ collapses under uncertainty of 5\%. It confirms that the velocity profiles at the inlet were measured using a sufficiently short wire length to avoid small-scale energy attenuation.

\section{Results} %%%%%%%%%%%%%%%%%%

\subsection{Performance assessment of the method.}

\begin{figure}
	\begin{center}
	\includegraphics*[width=0.8\linewidth]{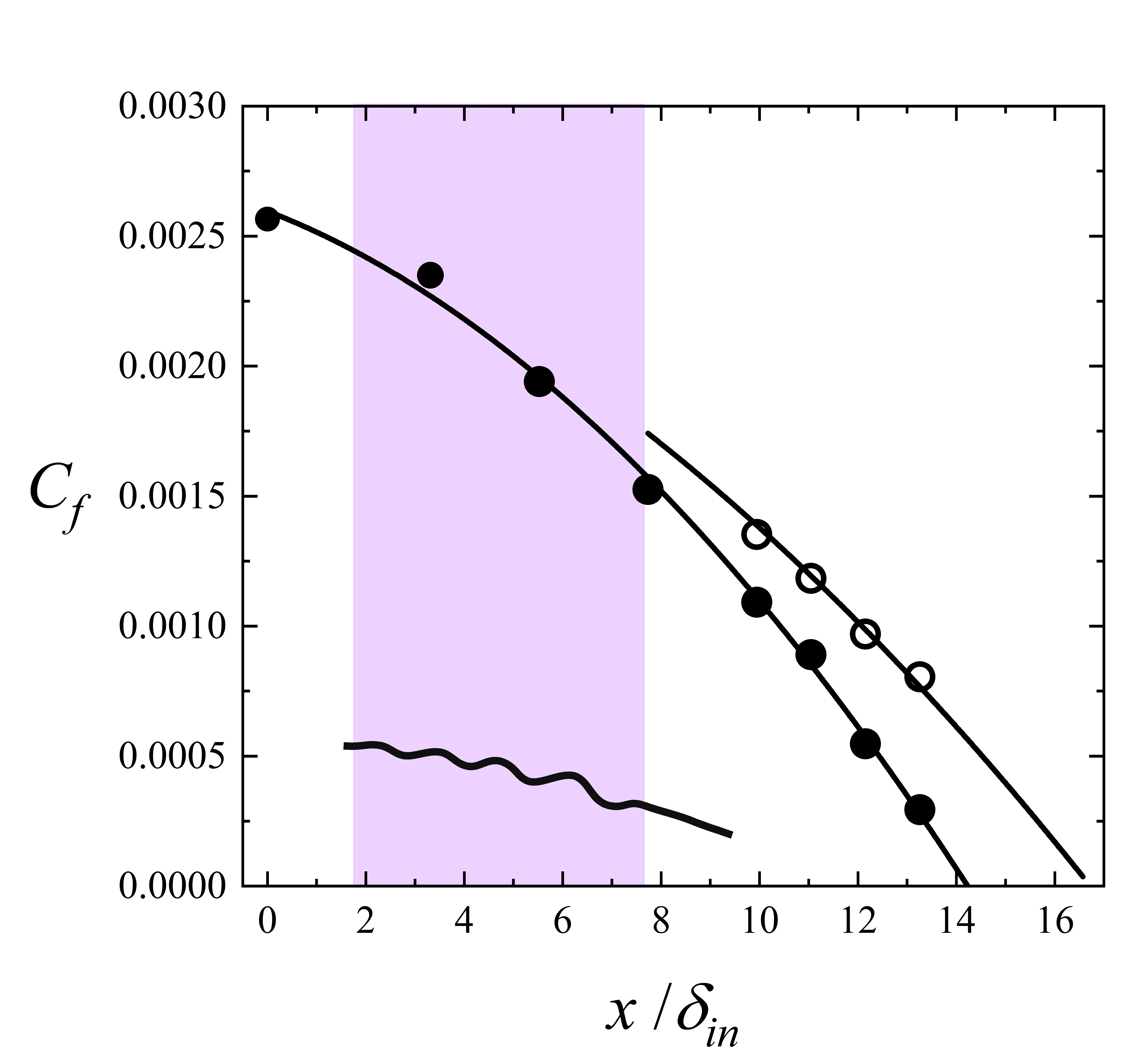}
	\caption{\label{fig_CfRe} Comparison of skin-friction coefficient distributions. The open symbol corresponds to the wavy wall, while the dark symbol corresponds to the unmodified wing surface.}
	\end{center}
\end{figure}

In Figure~\ref{fig_CfRe}, the skin friction coefficient $C_f = 2u_\tau^2U_e^{-2}$ for the wavy surface (open symbols) is compared to the unmodified wing surface $C_f$ (dark symbols), where a second-order polynomial extrapolated the distributions to predict the separation point. The streamwise distance from the inlet $x$ was reduced by the boundary layer thickness at the inlet $\delta_{in}$. The selected purple area indicates the flow region where the waviness is located. Figure~\ref{fig_CfRe} illustrates a marked enhancement in the skin-friction coefficient for the wavy wall configuration, as denoted by the open symbols. This increase is accompanied by a pronounced downstream displacement of the turbulent boundary layer separation point, indicating the efficacy of the wavy geometry in delaying flow detachment.

For a more statistically confident estimate of the effect of waviness on the skin friction coefficient, the differences between the $C_f$ distributions were integrated from the end of the wavy wall to the flow separation location that occurs for the flow without surface modification. The integration area does not account for the tangential separation shift $\Delta x_{t,s}$. The percentage value of $\Delta C_f$ was scaled by $C_{f, in}$ to assess the real effectiveness of the method.

\begin{figure}
	\begin{center}
	\includegraphics*[width=0.8\linewidth]{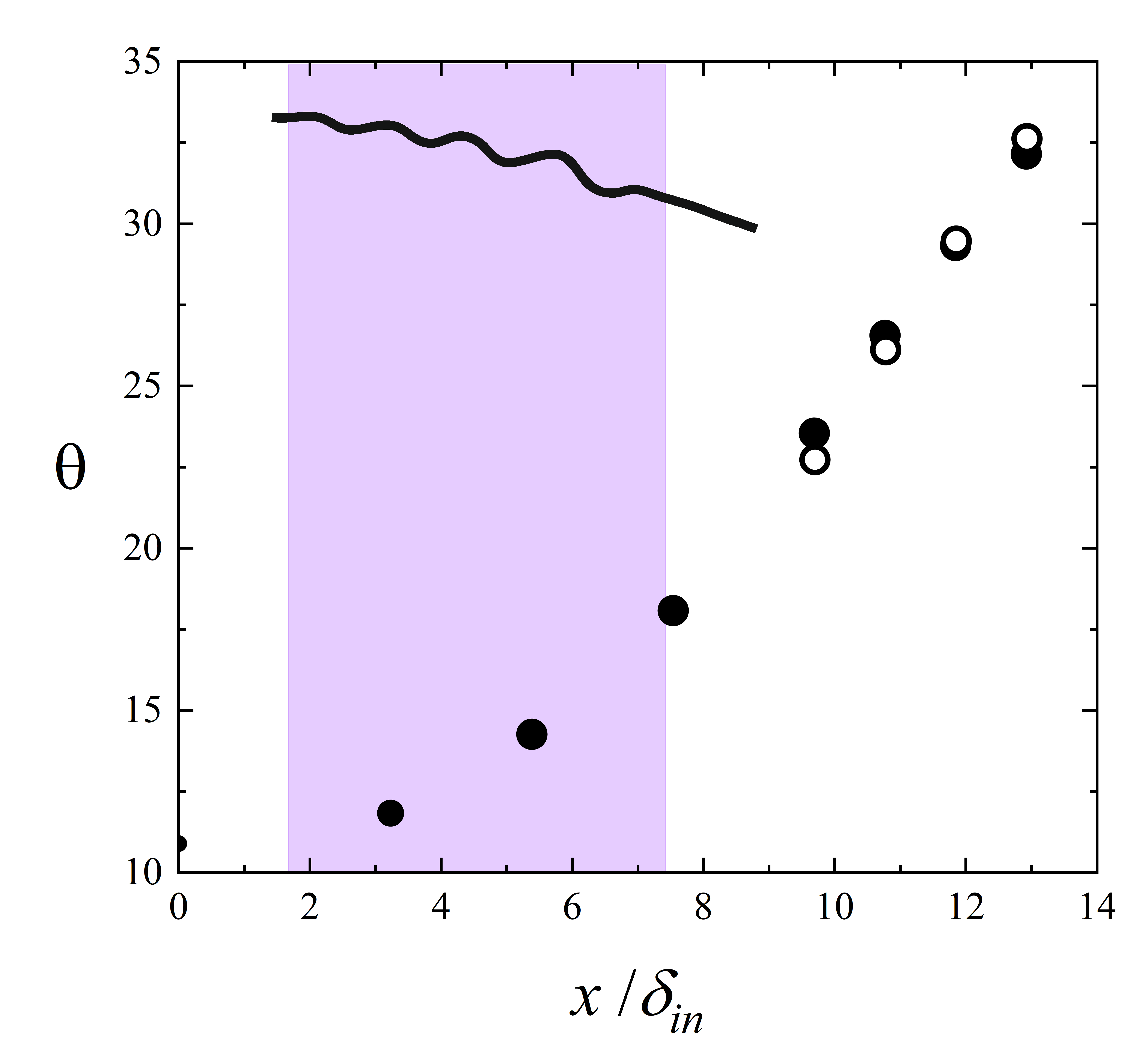}
	\caption{\label{fig_ThetaRe} Comparison of momentum loss thickness distributions. The open symbol is marked as a wavy wall case, while dark symbols correspond to the unmodified wing surface.
    }
	\end{center}
\end{figure}

The data in Table~\ref{tab:integr} show that the selected corrugation geometry produces a 13.4\%  increase in $C_f$, while reducing by value at the inlet $C_{f,in}$ (quantified in a way presented in Refs. \cite{drozdz2021effective, Drozdz2025}), which translates to an increase of 42.3\% in the integration area downstream of the wavy wall (quantified as in Ref. \cite{KAMINSKI2025135758}). 

The impact of pressure drag on overall airfoil efficiency remains a pertinent issue. In particular, Kamiński \emph{et al.} \cite{KAMINSKI2025135758} demonstrated that the overall pressure drag induced by the wavy wall is $0.00195$, which directly translates onto the increase in the drag coefficient $C_D$. However, this is an effect of the internal flow behaviour. For airfoils, the velocity deficit in the wake determines the drag coefficient. Due to the wavy wall mechanism preventing the extensive wake build-up, the airfoil's $C_D$ should be lower, as the $C_f$ gain induced by the enhanced sweeping motion is higher than the loss of $C_f$ due to the wavy wall pressure drag. This indicates that the pressure drag introduced by the wavy wall has a minor effect on the blade's drag coefficient.

\begin{table}
    \centering
        \caption{Integrated $C_f$ difference}
    \label{tab:integr}
    \begin{tabular}{cccccc}
        \hline
         Int. area [m] & $\Delta x_{t,s}$ [m] & $\Delta x_{t,s}\delta_{in}^{-1}$ [-] & $\Delta C_f$ [-]& $\Delta C_f$ [\%] & $\Delta C_fC_{f, in}^{-1}$ [\%]\\ 
        \hline
        0.7 - 1.3 & 0.21 & 2.32 & 0.000348 & 42.3& 13.5\\
        \hline
    \end{tabular}
\end{table}

To assess the reduction in velocity deficit in the wake of the airfoil, it is instructive to examine the evolution of the momentum-loss thickness, $\theta$. An effective separation control strategy should not merely redistribute momentum toward the wall but rather preserve or enhance the total momentum within the boundary layer. In this context, 
$\theta$ serves as a robust integral measure for quantifying the net momentum retention downstream of the control intervention. In Figure~\ref{fig_ThetaRe}, the distributions of $\theta$ for the wavy surface (open symbols) are compared with the distributions of $\theta$ as a function of $x\delta_{in}^{-1}$ on the unmodified wing surface (dark symbols). Symbols as in Figure~\ref{fig_CfRe}.  

\begin{figure*}
	\begin{center}
	\includegraphics*[width=0.7\linewidth]{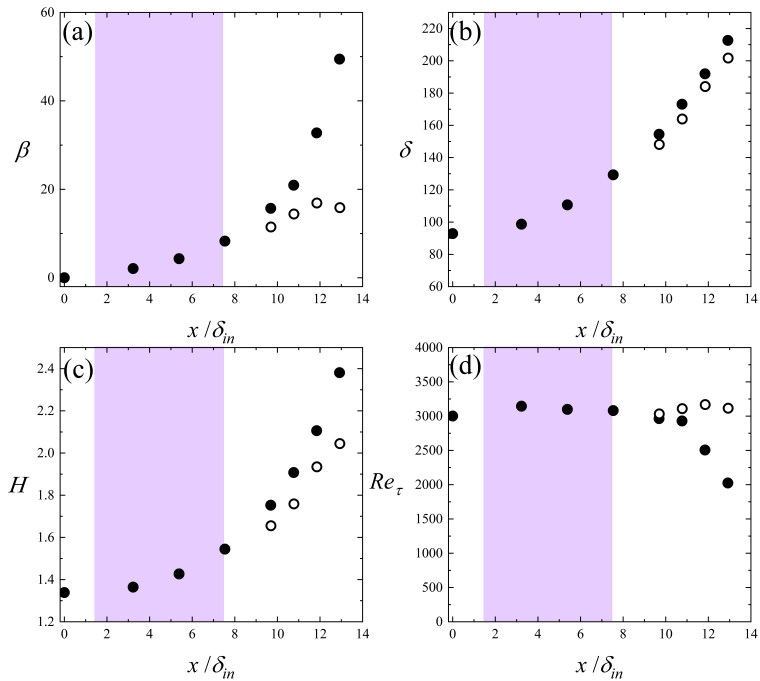}
	\caption{\label{fig:HdbRe} Comparison of $\beta$ a), $\delta$ b), $H$  c) and $Re_{\tau}$ d) downstream the flow.
}\end{center}
\end{figure*}

The results shown in Figure~\ref{fig_ThetaRe} indicate a maintenance of overall momentum by the wavy wall. The increase in the linear measure of the momentum-loss thickness observed in the final profile (at $x=1.2$ m) does not necessarily indicate that the wake width increases. The wake behind the airfoil is defined in the plane perpendicular to the main flow direction, resulting from the boundary layer developing on the suction and pressure sides of the airfoil. Since the boundary layer thickness for this location is approximately 5\% lower (see Figure~\ref{fig:HdbRe}) the velocity vector is more parallel to the surface, which indicates reduction of the wake.  Furthermore, the wake velocity deficit is affected by the interaction between boundary layers on the airfoil's upper and lower surfaces. The wavy shape decreases this mixing by increasing momentum close to the wall on the upper side, which helps balance the boundary layers. 

According to Atzori \emph{et al.} \cite{Atzori2020}, who studied suction and blowing applied on the suction side of the airfoil, an 11\% increase in the lift-to-drag ratio can be achieved by reducing wall-normal convection on the suction side using suction. The reduction in $\delta$ observed in Fig ~\ref{fig:HdbRe}~b) points to a decrease in wall-normal convection, which is consistent with the findings of Atzori \emph{et al.} \cite{Atzori2020}. This indicates that the baseline $C_d = 0.009$ of an airfoil at AoA of $5^\circ$ for $Re_c=10^7$ can be reduced by the wavy wall by limiting wall-normal convection that impacts pressure drag. If the surface geometry is suboptimally designed and causes local separation within the wave troughs, this mechanism is negated, and pressure drag begins to dominate, impairing aerodynamic performance. Although the exact increase in the lift coefficient $C_l$ remains undetermined, a tangential-to-the-wall shift in the flow separation point quantified as $\Delta x_{t,s}c^{-1} = 8.3\%$, where $c$ is the chord of the applied wing section, can be associated with a several percent increase in lift, which is considered a huge improvement in the aerodynamic performance of the airfoil at this Reynolds number. However, a critical methodological limitation is that the experimental domain terminates at $x/\delta_{in} \approx 13.3$, placing the extrapolated separation location approximately $x/\delta_{in} \approx 3.2$ downstream of the last measurement station. Given that the uncertainty in $ C_f$ for the last measurement is $ \approx 10\%$, extrapolating a fit based on a few downstream data points inherently amplifies the sensitivity to measurement error. This introduces an estimated uncertainty in the separation shift of $ 2\text{--}3\%$ of the chord ($\approx \pm 0.05\text{--}0.08,\text{m}$, corresponding to a relative uncertainty of $25\%$ in $\Delta x_{t,s}$. Consequently, while the qualitative efficacy of the wavy wall in suppressing separation is solidly evidenced by a statistically significant reduction in the shape factor (see Tab.~\ref{tab:Re}) and a $160\%$ increase in $C_f$ at $x/\delta_{in} \approx 13.3$, the reported quantitative shift ($\Delta x_{t,s} = 0.21\pm0.05 \text{m}$) should be interpreted as an estimate subject to substantial extrapolation margins rather than a definitive value.
The recent numerical investigation of Kamiński \emph{et al.} \cite{Kaminski2026}, performed at comparable Reynolds numbers but with a wavy wall that extends further downstream by one period, predicted exactly separation point location, and reveals a smaller tangential shift for both symmetrical and tilted wavy walls. 

\begin{table*}
  \centering
  \caption{Experimental TBL parameters for unmodified wing and wavy surface}
  \label{tab:Re}
  \begin{tabular}{cccccccccccc}

    $x\delta_{in}^{-1}$ & $x$ & $U_e$ & $\delta$ & $Re_{\theta}$ & $Re_{\tau}$ & $C_f$ & $\delta^*$ & $\theta$ & $H$ & $l^+$ \\
    \hline
\multicolumn{10}{c}{NACA4412 unmodified surface}	\\	

         3.31 & 0.3 & 14.3 & 98.7 & 10999 & 3145 & 0.00235 & 16.1 & 11.8 & 1.36 & 38.2 \\ 
        5.52 & 0.5 & 13.8 & 110.7 & 12816 & 3098 & 0.00194 & 20.4 & 14.3 & 1.43 & 33.6 \\ 
        7.73 & 0.7 & 13.3 & 129.3 & 15592 & 3081 & 0.00153 & 27.9 & 18.1 & 1.54 & 28.6 \\ 
        &&&&&&&&&&& \\
        9.94 & 0.9 & 12.7 & 154.5 & 19336 & 2964 & 0.00109 & 41.3 & 23.6 & 1.76 & 23.0 \\ 
        11.05 & 1.0 & 12.4 & 173.1 & 21310 & 2930 & 0.00089 & 50.7 & 26.6 & 1.91 & 20.3 \\ 
        12.15 & 1.1 & 12.3 & 191.9 & 23140 & 2506 & 0.00055 & 61.7 & 29.3 & 2.11 & 15.7 \\ 
        13.26 & 1.2 & 12.2 & 213.5 & 25269 & 2032 & 0.00029 & 76.7 & 32.2 & 2.38 & 11.4 \\

\hline																					
	\multicolumn{11}{c}{with wavy surface}	\\			
     9.94 & 0.9 & 12.6 & 148.0 & 17699 & 3032 & 0.00135 & 37.1 & 22.5 & 1.65 & 8.4 \\
        11.05 & 1.0 & 12.3 & 163.9 & 20349 & 3108 & 0.00118 & 46.0 & 26.1 & 1.76 & 7.8 \\ 
        12.15 & 1.1 & 12.5 & 183.9 & 23035 & 3165 & 0.00097 & 56.9 & 29.5 & 1.87 & 7.1 \\ 
        13.26 & 1.2 & 12.2 & 201.7 & 25112 & 3115 & 0.00081 & 66.9 & 32.6 & 2.05 & 6.3 \\

\hline																					

  \end{tabular}
\end{table*}
The values of $Re_\theta$ and other important TBL parameters of eleven velocity profiles determined in the flow over the wing section are shown in Table~\ref{tab:Re}. It can be concluded that the optimal configuration causes the local $Re_\theta$ to decrease when using a wavy wall. Other parameters indicating the effectiveness of the proposed flow control method are presented in Figure~\ref{fig:HdbRe}. The primary parameter most affected is $\beta$ as it includes the wall shear stress that is strongly influenced (see Figure~\ref {fig:HdbRe}~a). The distributions show a considerable decrease in the value of $\beta$ as a result of the impact of wavy wall. The increase in strain rate, which is proportional to the wall shear stress, is also reflected in the reduction in the thickness of the TBL (see Figure~\ref {fig:HdbRe}~b) and in the shape factor $H$ (see Figure~\ref {fig:HdbRe}~c). The most important observation concerns $Re_{\tau}$ distributions. As can be seen in Figure~\ref {fig:HdbRe}~d) this distribution downstream of WW, unlike in the reference case, is practically constant. This means that the overall balance between large-scale fluid motion and near-wall viscous forces remains almost constant.

\begin{figure*}
	\begin{center}
	\includegraphics*[width=0.8\linewidth]{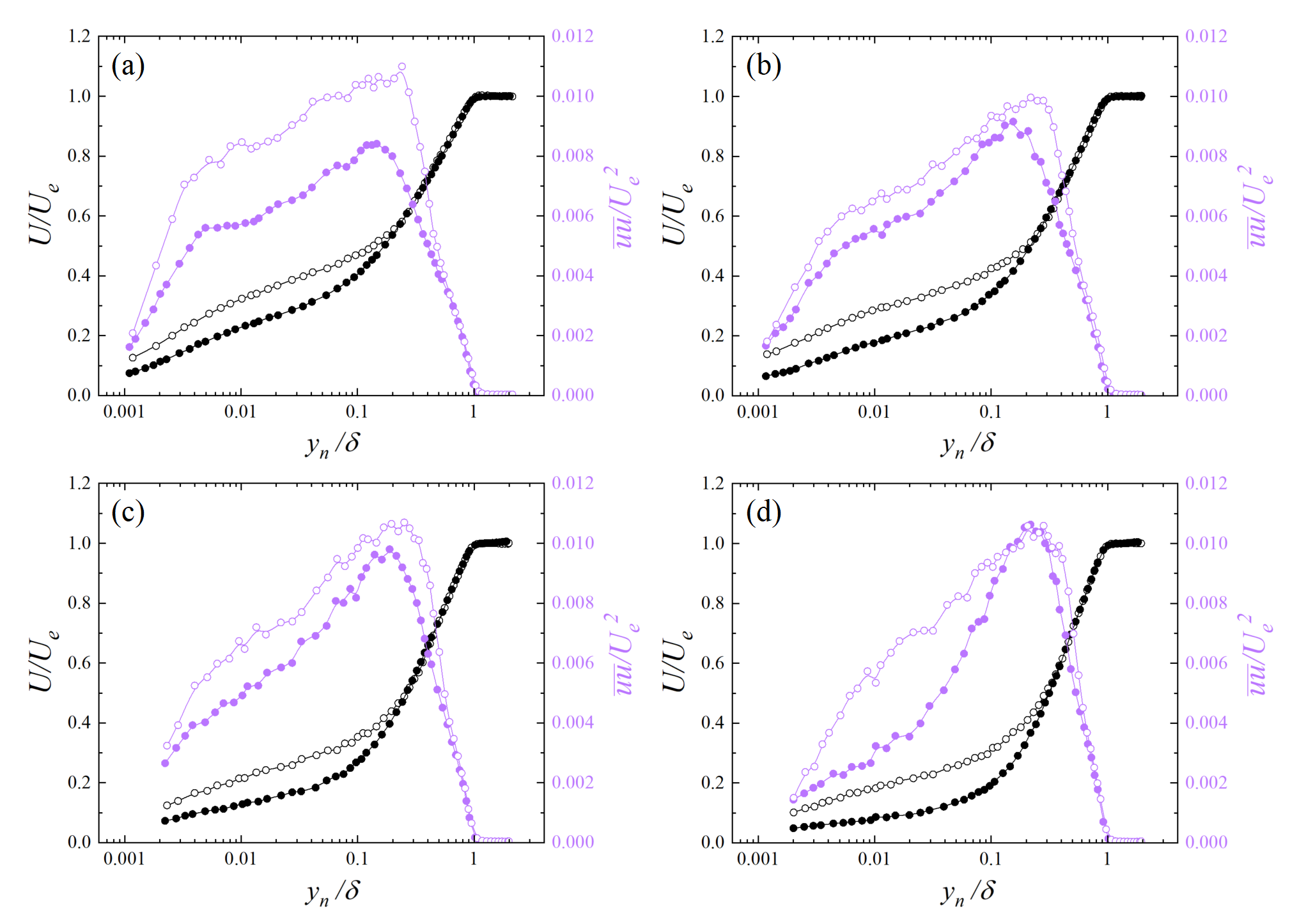}
	\caption{\label{fig_profiles} Comparison of the mean velocity and velocity variance components: $x=0.9$ m a), $x=1.0$ m b), $x=1.1$ m c) and $x=1.2$ m d).
}\end{center}
\end{figure*}

To confirm the above statement and to gain a more detailed insight into the flow field modified by the wavy wall, the wall-normal profiles of mean velocity and streamwise Reynolds stress downstream of WW are shown in Figure ~\ref{fig_profiles}. Note that the profiles are normalised by the edge velocity $U_e$. The streamwise evolution of these profiles is shown at $x=0.9$ m, $x=1.0$ m, $x=1.1$ m, and $x=1.2$ m at the surface level. The black symbols correspond to the profiles measured on the unmodified wing surface, while the open symbols represent the profiles for the WW case.

As was concluded by Hudson \emph{et al.} \cite{Hudson1996}, the flow over the wavy wall can be divided into an outer region, where the effects of the wavy surface are small, and an inner region, where the effects are significant. Fig.~\ref{fig_profiles} shows that the boundary layer developing behind the wavy wall is also affected only near the wall. The mean velocity increases due to the wavy wall. A direct comparison between $\overline{uu}U_e^{-2}$ profiles indicates that the wavy surface is responsible for flattening the outer maximum. The flattening rate increases with increasing $x$. This effect can be attributed to the change in the flow history of the pressure gradient, as the distribution $\beta$ (see Figure ~\ref{fig:HdbRe}) is weaker for a wavy wall, causing the outer maximum to shift away from the wall \cite{bobke_vinuesa_örlü_schlatter_2017}. Figure~\ref{fig_profiles}a in $x=0.9$ m shows that $\overline{uu}U_e^{-2}$ is leveraged by the wavy wall and decays faster than on the unmodified surface. It can be concluded that, when we deal with the effective flow control strategy according to Atzori \emph{et al.} \cite{Atzori2020}, the reduced outer-layer fluctuations translate into a reduced wake behind the airfoil. There is also a noticeable increase in $\overline{uu}U_e^{-2}$ distributions in the near-wall region. It is far from the unmodified case in the last profile (Fig.~\ref{fig_profiles}~d), where the level of $\overline{uu}U_e^{-2}$ is low as the flow is close to separation.

\begin{figure*}
	\begin{center}
	\includegraphics*[width=0.618\linewidth]{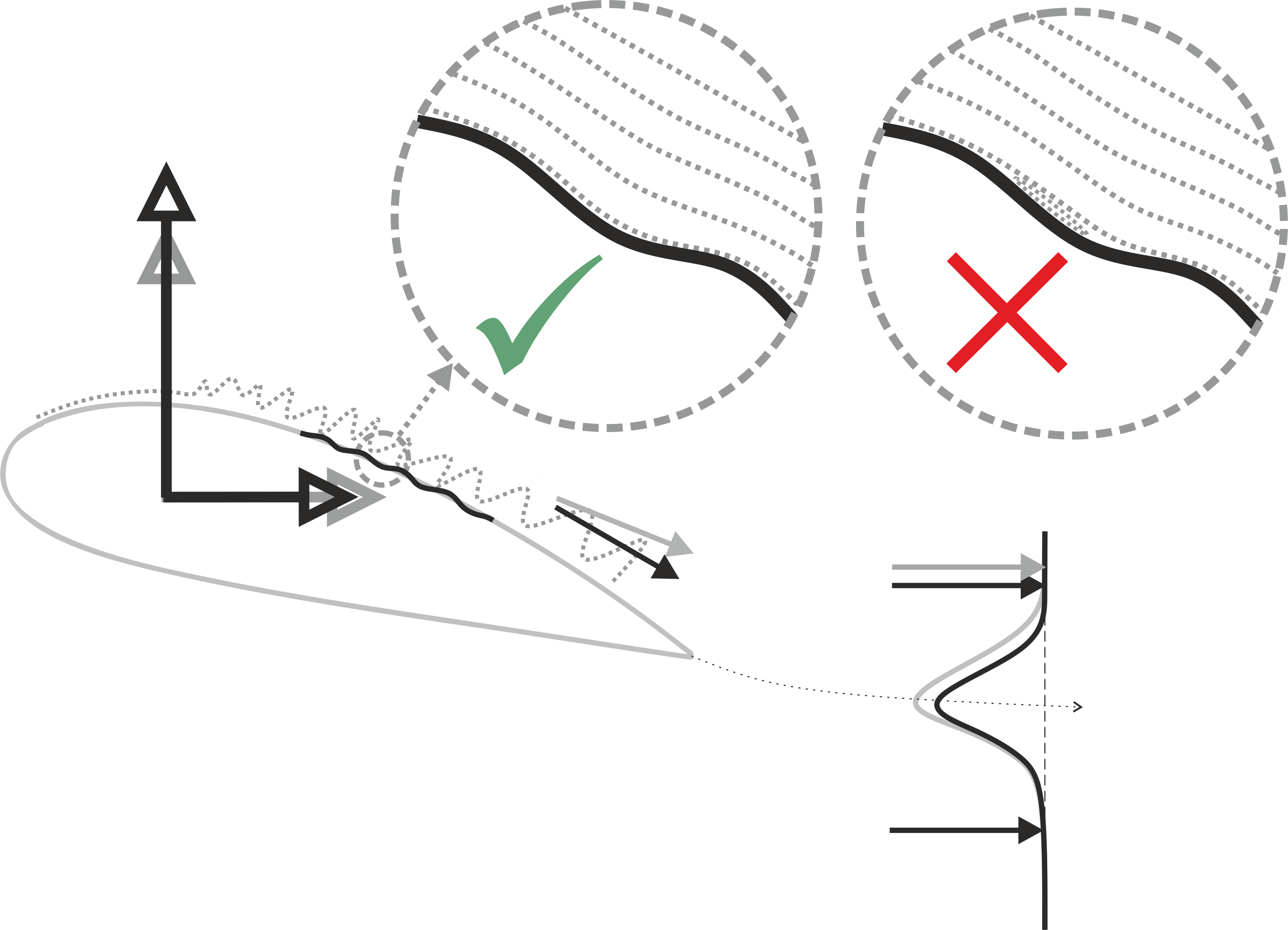}
	\caption{\label{fig:GA} Schematic mechanism of the wavy wall at high Reynolds numbers. The dark colour corresponds to the characteristics (lift force, drag force, and wake) of a wavy wall with a flow at the verge of separation in troughs.
}\end{center}
\end{figure*}

\subsection{New insights into physical effects due to wavy wall}

\begin{figure}
	\begin{center}
	\begin{minipage}[c]{0.58\linewidth}
		\centering
		\includegraphics*[width=\linewidth]{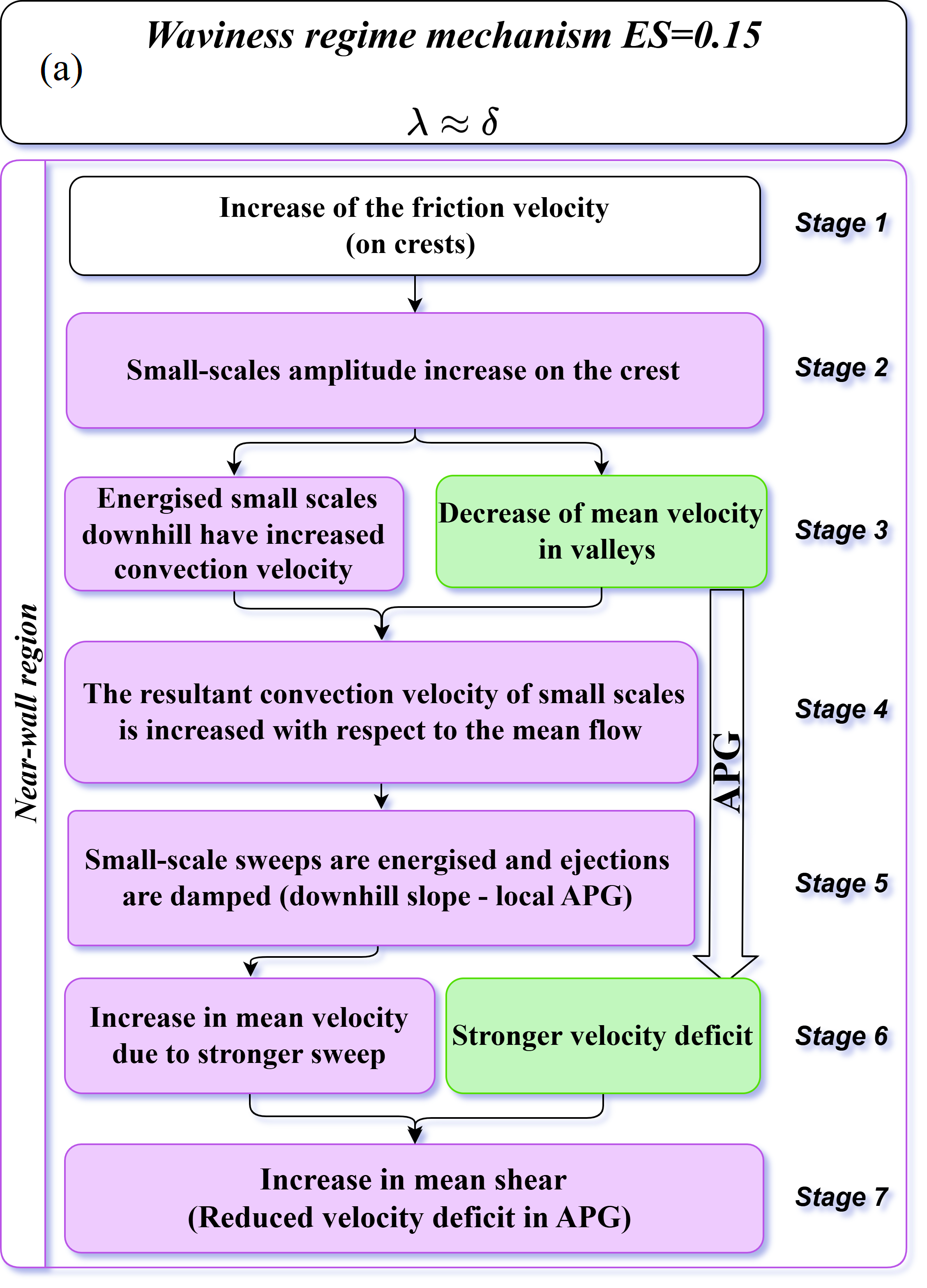}
	\end{minipage}%
	\hspace{0.03\textwidth}%
	\begin{minipage}[c]{0.38\linewidth}
		\centering
		\includegraphics*[width=\linewidth]{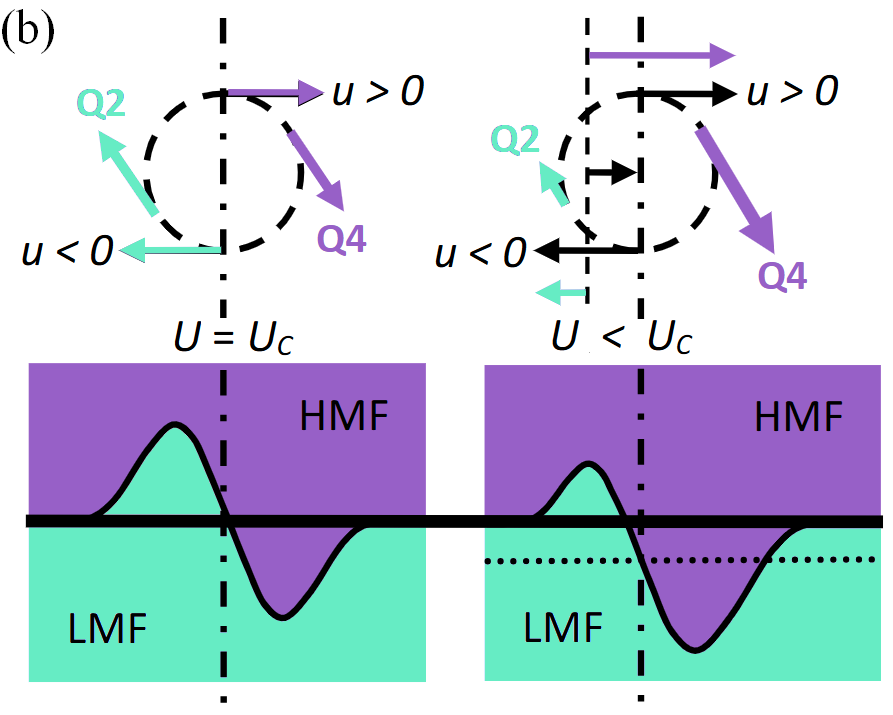}
	\end{minipage}
	\caption{\label{fig:Q4} Diagram of sweeping motion increase in spanwise vortex steaming from the crest of wavy wall at high Reynolds numbers (a). Schematics of sweep increase sourced from Dróżdż \cite{Drozdz2014c} (b).
}\end{center}
\end{figure}

Figure \ref{fig:GA} presents the mechanism induced by the wavy-wall, provided that there is no flow separation in the trough at a high Reynolds number. Under these conditions, a strong shear layer forms upstream of the wave peak and promotes the largest difference between the mean flow and the convection of small-scale spanwise vorticity on the downhill slope of the waves. The downhill slope generates the local APG flow, which, on average, remains attached when dealing with a wavy wall in the waviness regime ($\lambda \approx \delta$) \cite{nugroho_monty_utama_ganapathisubramani_hutchins_2021}, increasing the convective velocity and, accordingly, the sweeping motion, as described in the 7-stage diagram in Fig.~\ref{fig:Q4}~a). These flow conditions increase the sweeping motion, specifically underlying the concept illustrated in Figure~\ref{fig:Q4}~b) taken from Dróżdż \cite{Drozdz2014c}. Figure~\ref{fig:Q4}~b) shows the strength of quadrant events produced by a spanwise vortex for two cases: determined by different convection velocities. The upper schematics present a small-scale vortical structure at different convection velocities with respect to the mean flow and show how its velocity influences ejection (Q2) and sweep (Q4) events. Below the schematics of the vortical structures, changes in momentum-transport areas are shown. For simplicity, it was assumed that high-momentum fluid (HMF) is present above the vortex center, while low-momentum fluid (LMF) is present below it. The case on the left in Figure~\ref{fig:Q4}~b) shows that when the  $U_C$ and $U$ are equal (occurring in ZPG), the energies of the sweep and ejection events induced by the vortex are also equal (except for the region below the inner peak in streamwise Reynolds stresses). The LMF occupies the same area as the HMF because the same amount of fluid is ejected as the fluid is swept. This results in no change in the total momentum in the area. In the second case (favourable in APG) the convection velocity is higher than the mean velocity; the stronger, by the difference between $U_C$ and $U$, positive streamwise event $u$ is transferred by viscosity in the wall-normal direction and reinforces Q4 event. Meanwhile, the same mechanism also reduces Q2. The HMF area is greater than the LMF area because more fluid is swept than is ejected, resulting in an increase in the total momentum in the area.
In summary, the momentum near the wall is higher in APG flows because the energy is transferred from the ejection to sweep. The mechanism by which a wavy wall delays turbulent boundary-layer separation in the APG region is then clarified. The key factor is the slope of the hill downstream, where sweep events are enhanced. These findings advance our understanding of wall-normal momentum transfer in turbulent boundary layers in APG, owing to increased convective velocity. The number of small-scale vortices increases in high-speed, large-scale regions and decreases in low-speed regions, resulting in an imbalance in the populations of ejections and sweeps. The resulting increase in convective velocity influences the quadrant events that drive wall-normal momentum transfer.

This mechanism also reduces wall-normal convection as evidenced by lower $\delta$ (Fig.\ref{fig:HdbRe}~b). It was confirmed by Atzori \emph{et al.} \cite{Atzori2020}, that the reduced wall-normal convection promotes a momentum redistribution to the near-wall region of the airfoil suction surface, directly decreasing the deficit in the wake of the airfoil.
In the other case, when separation occurs, the WW generates a large pressure drag and accompanying skin-friction loss; therefore, the mechanism of the increase in sweep events(see Fig.~\ref{fig:Q4}~b) is ineffective as the convection is equal to the mean flow.

The physical mechanism of an increase in momentum by a wavy wall in the aspect of small-scale activity was already discussed in Dróżdż \emph{et al.} \cite{Drozdz2025}. The key condition to promote a momentum-increasing wavy wall is high energy of small scales over the wavy surface that is flown without separation in the trough. This condition is restricted by large-scale motions, as indicated by an outer peak appearance that exceeds the inner peak, signalling a shift in flow-driven energy from small to large scales. When small-scale activity remains elevated, it signifies enhanced momentum transport induced by the wavy wall. The decreasing fraction of small scales below a certain level within the APG flow region indicates a loss of efficiency in the wavy wall, which limits its usability. The limiting value of $\beta$ depends on the effect of the flow history, as it is known from the literature that the appearance of the outer peak in $\overline{uu}$ decays with the strength of APG. When the streamwise evolution of $\beta$ is slow, the outer peak appears earlier and flow-driven energy changes from small to large scales. When the limit is defined correctly, the density of small‑scale eddies \citep{Herpin2013} above the wavy wall is sufficiently high; the small scale drives the near-wall flow. As a result, the wavy wall promotes the production of small‑scale turbulence downstream, enhancing mixing and preventing flow separation.

However, this beneficial behaviour occurs only when the valley region remains free of separation (see Fig.~\ref{fig:GA}). These conditions define the appropriate amplitude and effective slope for each wave period, but only for a specific friction Reynolds number, because the mechanism relies on a sufficiently dense population of small‑scale eddies.

In APG flows with a pronounced outer peak in the streamwise Reynolds stress, large‑scale motions dominate and the friction Reynolds number decreases. Under these conditions, the density of small‑scale eddies becomes too low for the wavy‑wall mechanism to operate effectively, limiting the enhancement of turbulent mixing. Any flow separation occurring within the wavy‑wall valley further increases pressure drag and suppresses the intended skin‑friction benefit. This degradation happens even when the amplitude and period were originally tuned for high‑Reynolds‑number conditions with abundant small‑scale turbulence.

\begin{figure*}
	\begin{center}
	\includegraphics*[width=0.8\linewidth]{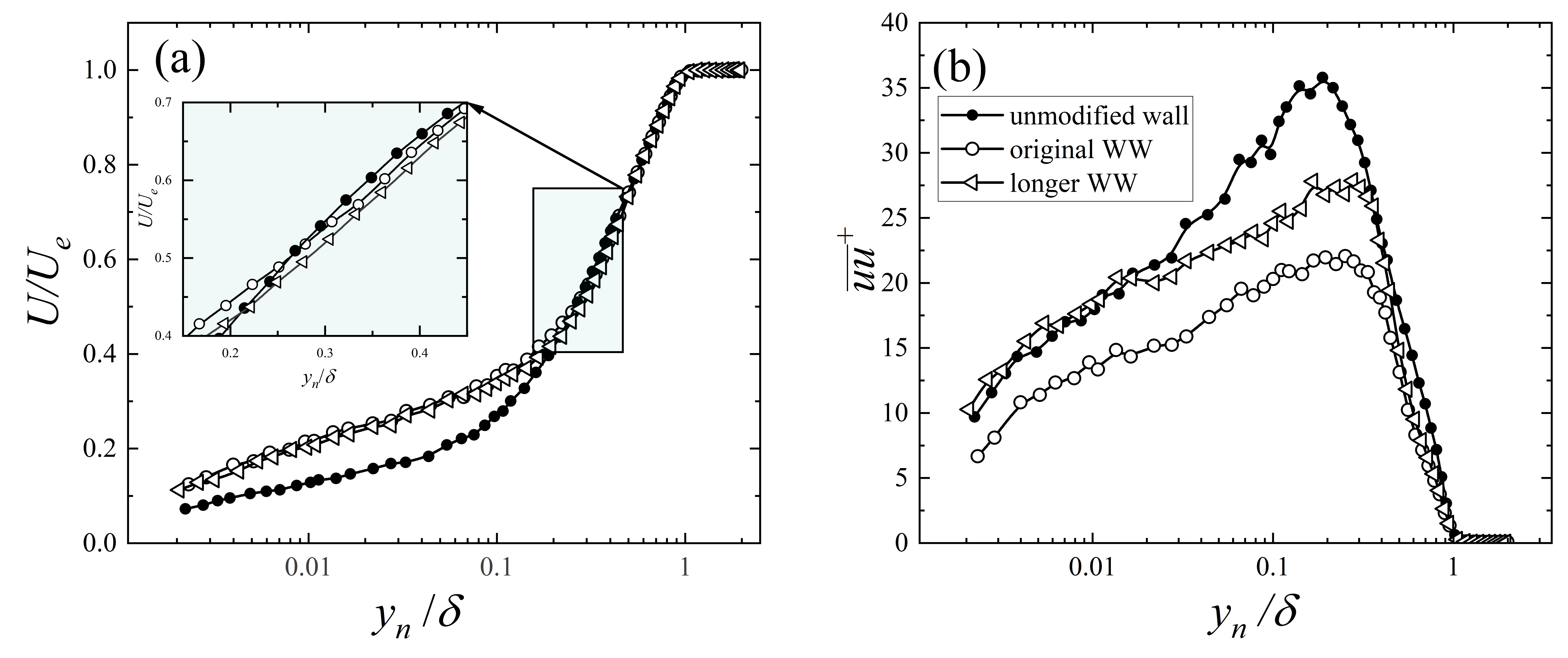}
	\caption{\label{fig_profs} Comparison of the $U/U_e$ (a) and $\overline{uu}^+$ (b) components at $x=1.1$ m. Symbols as in Figure \ref{fig_profiles}, while results with longer wavy wall are marked with left triangles.
}\end{center}
\end{figure*}

To illustrate this efficiency-limiting mechanism, we consider a configuration in which the wavy wall is extended further downstream. The results presented in Figure~\ref {fig_profs} were obtained for two cases, i.e. for a wavy wall ending at $\beta \approx 8$ (see Figure~\ref {fig:HdbRe}~a) and for the case with extended wavy wall by one period of 0.135 m, which gives the end of WW at $\beta \approx 12$. The comparison of mean velocity profiles and viscous scaled Reynolds stress is presented in Figure~\ref {fig_profs}~a) and Figure~\ref {fig_profs}~b), respectively. Additional profiles were taken at the same location on the unmodified surface for comparison. 

Figure \ref{fig_profs}~a) reveals (see blow-up in the figure) that the velocity deficit for longer WW is greater up to $y_n\delta^{-1} \approx 50\%$ of the thickness of the TBL, while Figure \ref{fig_profs}~b) shows that the viscous-scaled Reynolds stresses $\overline{uu}^+$ decrease at the outer peak for both cases of WW, similar to the work of Atzori \emph{et al.}\cite{Atzori2020} when suction was implemented on the suction side. Interestingly, for the shorter WW, $\overline{uu}^+$ is also reduced near the wall, indicating that the method more effectively transports momentum to the wall while simultaneously reducing the overall turbulence level.

\begin{figure*}
	\begin{center}
	  
    \includegraphics*[width=0.8\linewidth]{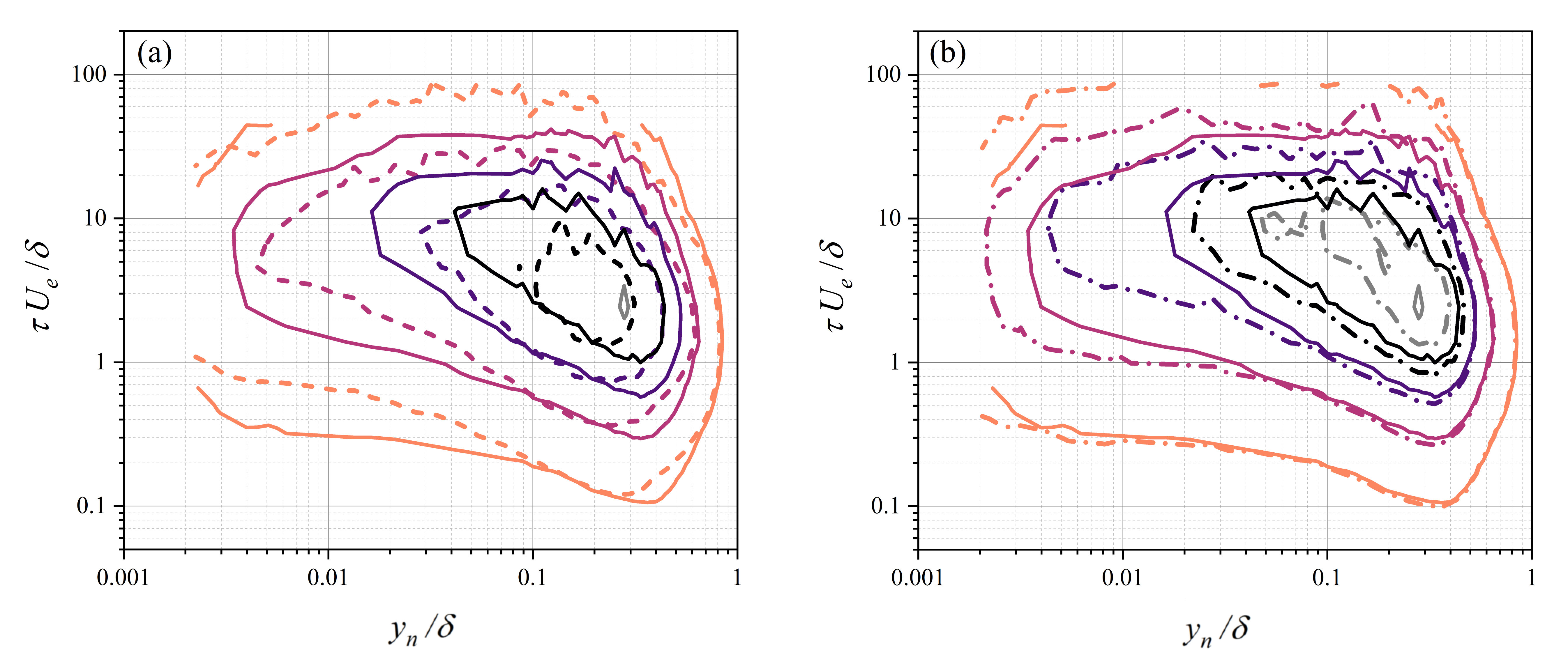}
    
	\caption{\label{fig:spectra} Premultiplied energy spectra comparison  at location $x = 1.1$ m. Solid lines original WW and dashed contours wing profile surface (a) and dashed-dotted contours WW extended by one period (b). Increments marked by colours between iso-contour levels equal 0.0005.}
     
    \end{center}
\end{figure*}

A higher velocity deficit indicates a lower increase in the skin-friction coefficient despite a clear increase in energy troughout the entire boundary layer thickness. To explain the role of small and large scales fractional energy contribution in the flow, the iso-contours of the reduced wavelet energy spectra $E_{uu}U_e^{-2}$ (equivalent to the pre-multiplied energy spectra) are presented as a function of the normalised time scale $\tau U_e\delta^{-1}$ and the normalised wall distance $y\delta^{-1}$ in Figure~\ref {fig:spectra} for the same location for these cases.
In Figure \ref{fig:spectra}~a), an increase in large-scale energy across the TBL thickness and a small-scale energy increase near the wall due to original WW (solid line) is observed. A similar enhancement of small-scale energy is observed in a longer wavy wall (see dashed-dot line in Figure \ref{fig:spectra}~b); however, the higher increase in large-scale energy occurs for that case. It indicates that a too-high Reynolds stress level (see Figure \ref{fig_profs}) further downstream of the wavy wall is a result of the large-scale energy increase caused by a WW extending deep into the APG region. It indicates that, the reduction in large-scale energy ensures that the argument of Atzori \emph{et al.} \cite{Atzori2020} is correct: that is, a net reduction in drag — manifested by delayed separation, decreased wake strength, and lower drag coefficient — is possible through the decrease in large-scale energy accompanied by the increase in the skin-friction coefficient on the suction side of the airfoil. Conversely, stronger large-scale energy intensifies wake mixing, weakening flow control. Although flow remains improved over the smooth baseline, the extended geometry reduces peak efficiency and net momentum gain compared to the optimal 4-period setup.

It appears that the mechanism of the outer peak decay is related to the small-scale energy conservation along the flow. The increase in small-scale energy by the Wavy wall may lead to the breakup of large-scale motions in a similar way as the large-eddy break-up (LEBu) device, which is in fact a small-scale turbulence production device as was shown in Ref. \cite{Chin2017}.

\begin{figure}
	\begin{center}
	  
    \includegraphics*[width=0.62\linewidth]{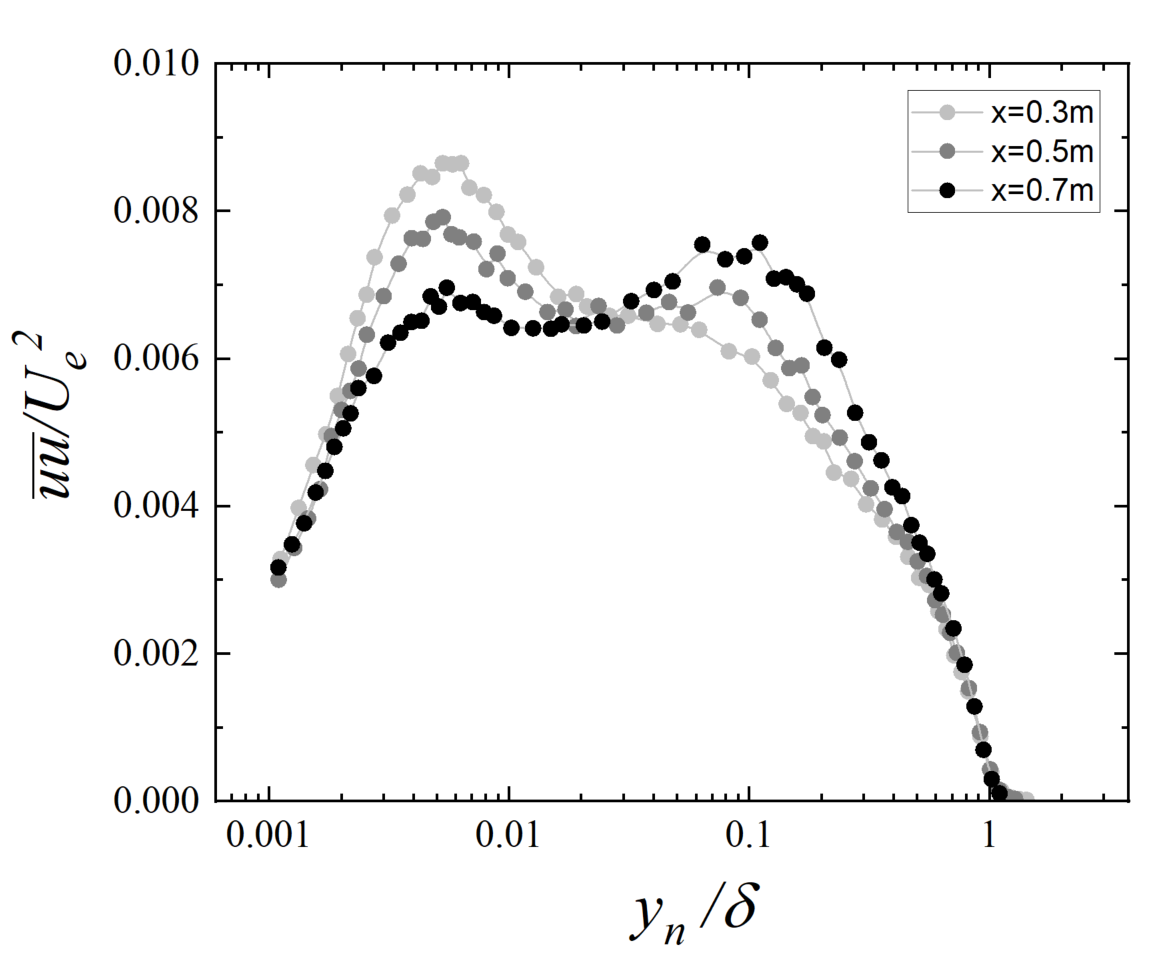}
    
	\caption{\label{fig:uu_flat} Normalised variance distribution over the measurement domain (0.3 to 0.7 m) }
     
    \end{center}
\end{figure}

The effect discussed above explains specific differences between the experimental and numerical results in assessing how the wavy wall affects the flow structures. The numerical simulations of Kamiński \emph{et al.} \cite{Kaminski2026} report a distinct increase in large-scale energy observed in the velocity premultiplied spectra distributions attributed to the WW effect. The small-scale energy increase was observed only for the tilted wavy wall in Ref. \cite{Kaminski2026}.
The difference is because, in the numerical study \cite{Kaminski2025,Kaminski2026}, the WW extended to $ \beta \approx 10 $, which was accepted according to the first studies of Dróżdż \emph{et al.} \cite{drozdz2021effective}. In the current experimental setup, the WW terminates earlier than in Ref. \cite{Kaminski2026}, at $\beta \approx 8$, which is consistent with the earlier outer-peak enhancement (see profile $x = 0.7$ m in Fig. \ref{fig:uu_flat}). Furthermore, the additional period of WW exhibited in the experimental study an amplitude exceeding $A^+=170$, resulting in decreased control efficiency and substantial large-scale production.

In summary, excessively long WW and excessively high amplitudes may trigger an increase in large-scale motion, reducing the efficiency of the WW mechanism. These findings emphasise that an optimally designed WW enhances the population of small-scale structures without contributing to large-scale energy production. Although the first studies Dróżdż \emph{et al.} \cite{drozdz2021effective} suggested terminating WW at $ \beta \approx 10$, however, as shown in subsequent studies \cite{Drozdz2025}, this threshold is not strictly governed by the value of $ \beta $. Instead, the decisive criterion is the emergence of a distinct outer peak in Reynolds stress that exceeds the inner peak:
$ \max \left( \overline{uu}_{\text{outer}} \right) > \max \left( \overline{uu}_{\text{inner}} \right)$.  
The recommendation also applies to cases with different flow histories.

According to this criterion, for the present case the location $x=0.7$ m marks the upstream limit for effective wavy‑wall application (see Figure \ref{fig:uu_flat}). Upstream of this point, the inner peak—and thus small‑scale activity—still dominates near‑wall turbulence. Crossing this limit shifts the dominant production mechanism toward outer‑layer motions, reducing the ability of the wavy wall to control separation. Beyond $x=0.7$ m, the wavy‑wall influence weakens, and the momentum‑loss thickness increases. 

Therefore, an optimal WW design should extent small-scale universality further downstream by minimising excessive Reynolds stress in the outer layer. By tracking the evolution of turbulent structures, particularly when the outer peak surpasses the inner peak, the streamwise extent of the WW can be fine-tuned to maximise control efficiency without inducing large-scale motions. Conversely, if this limit is exceeded, large-scale motions become dominant, suppressing the enhancement of small-scale propagation speed and resulting in higher value of the momentum-loss thickness. 

\section{Conclusions and discussions} %%%%%%%%%%%%%%%%%%

A detailed experimental study shows that the addition of a wavy wall to a convex surface, in the specified location, may constitute a previously unexamined passive method to alter aerodynamic behaviour. It provides a new approach to delay flow separation and improve airfoil performance at high Reynolds numbers.

The presence of the wavy surface leads to a substantial rise in the skin-friction coefficient. Comparative results show that the selected corrugation shape may produce an increase of up to 42.3\% over the integration region extending from the end of the wavy wall to the location where separation would begin on a smooth, unmodified surface.
A significant outcome is the tangential shift in the flow separation point by 8.3\% of the chord, which is roughly associated with a several percent increase in lift for an airfoil at the given Reynolds number. This represents a substantial improvement in aerodynamic performance, primarily due to sustained momentum loss thickness and a few percent reduction in turbulent boundary layer thickness, indicating that the wall-normal convection is decreased due to an increase of streamwise convection caused by the wavy wall.
Mean velocity profiles show a notable increase in the inner part, along with a noticeable rise in Reynolds stress distributions and flattening of the outer maximum due to waviness occupying the near-wall region. Furthermore, an increase in strain rate also contributes to a reduction in the shape factor $H$. At the same time, the value of $Re_{\tau}$ remains at a constant level further downstream of the wavy wall, contrasting with observations on the unmodified section of the wing.

The comprehensive analysis of the presented research also sheds light on the mechanisms by which wavy surfaces influence aerodynamics, which are now better understood and emphasise the significant role of small-scale turbulence.
The fundamental feature driving the increase in momentum towards the wall is small-scale turbulence activity. The wavy wall works effectively when a high concentration of small-scale eddies is present above its surface. The underlying physical mechanism relies on the increased propagation speed of small-scale spanwise vortices relative to the mean velocity ($U_c > U$) over the downhill wave slope. This velocity differential transfers positive streamwise fluctuations ($u > 0$) toward the wall through viscous effects, thereby reinforcing Q4 sweeping events ($u > 0, v < 0$) while suppressing Q2 ejections ($u < 0, v > 0$). Consequently, the high-momentum fluid (HMF) domain predominates over the low-momentum fluid (LMF) domain, driving net momentum transport toward the wall, provided that the wave troughs remain unseparated.

However, the wavy wall's efficacy is critically bound. Its benefits diminish at high $\beta$ values, where large-scale motions become dominant. The presented results indicate that excessive wavy-wall length can trigger a detrimental increase in large-scale motion, reducing WW efficiency at enhancing momentum-increasing small-scale sweep events. The optimal termination criterion for a wavy wall is not a fixed $\beta$ value, as previously thought, but rather flow history and, related to it, the emergence of a distinct outer peak in the Reynolds stress that surpasses the inner peak. Beyond this point, the wavy wall's ability to control flow separation diminishes, and the momentum-loss thickness increases. Excessive wavy wall amplitude will have a similar effect.
This reinforces the conclusion that maintaining a high small-scale eddy density is the primary determinant of wavy wall effectiveness and superior momentum transport. It appears that small scales not only increase turbulent mixing but also increase the streamwise decay of large-scale motions in the outer region, possibly reducing the airfoil's wake.

In essence, the wavy surface, when precisely tuned, acts as a sophisticated flow control mechanism, effectively increasing skin friction and postponing flow separation, thus enhancing lift without a detrimental impact on overall airfoil drag, through favourable modifications to the boundary layer structure and momentum management. 

Future studies will investigate the enhanced performance achieved by introducing a non-harmonic wave type to a curved surface resembling a blade surface.

%%\backsection[Acknowledgements]{The investigation was supported by National Science Centre under Grant No. DEC-2020/39/B/ST8/01449.}

Data availability statement - The data that support the findings of this study are openly available in [RepOD] at https://doi:10.18150/6BIJKK,

\begin{acknowledgments}
The investigation was supported by National Science Centre under Grant No. DEC-2020/39/B/ST8/01449.
\end{acknowledgments}

%{Artur Dróżdż: Writing – original draft, Writing – review and editing, Visualisation, Validation, Supervision, Software, Methodology, Investigation, Formal analysis, Data curation, Conceptualisation. Mathias Romańczyk: Methodology, Investigation, Visualisation, Validation. Witold Elsner: Writing – review and editing, Supervision, Resources, Project administration, Investigation, Funding acquisition, Formal analysis, Conceptualisation, Validation.}

\appendix

\section{Attenuation }

\begin{figure}
	\begin{center}
	\includegraphics*[width=0.62\linewidth]{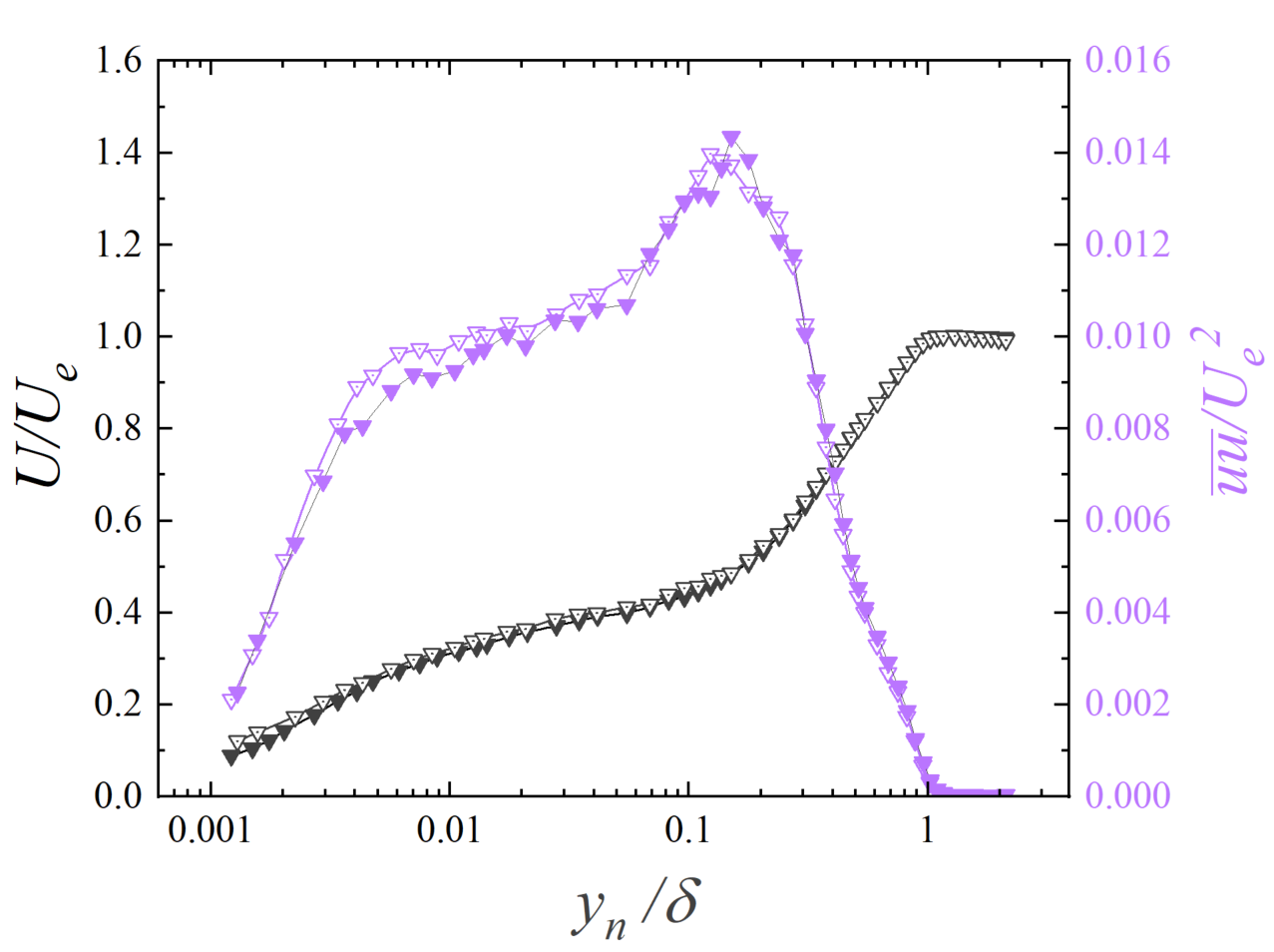}
	\caption{\label{fig:HWSA} Comparison between two types of hot-wire probes of the mean velocity and velocity variance components at $x=0.9$ m of the longer wavy wall. The open triangle corresponds to the gold-plated hot-wire probe 55P05. The dark triangle corresponds to modified hot wire 55P31 measurements.
}\end{center}
\end{figure}

In the appendix, the effects of hot-wire spatial averaging are discussed. According to Dróżdż \emph{et al.} \cite{Drozdz2024meas}, the hot-wire spatial averaging in APG flows in the inner and outer peaks in streamwise Reynolds stress $\overline{uu}$ can be observed up to $\beta =20$ due to a too long sensing length of the hot-wire. Above that value of $\beta$, the length of the wire of 100 viscous units gives an attenuation lower than the uncertainty of 5\%. The critical location for analysed data is location $x=0.9$ m for $\beta \approx 15$, where $l^+ = 23$ (see Table \ref{tab:Re}). To show the estimation reliability of the $\overline{uu}$ Reynolds stresses for the two probes, gold-plated and modified, were compared in Figure \ref{fig:HWSA} for the case with a longer wavy wall that produces slightly higher wall shear stress than the unmodified surface. The results reveal a discrepancy between profiles within the uncertainty range.

In the analysis, the crucial location is $x=0.7$ m at $\beta \approx 8$ and $l^+ = 29$, where the outer peak dominates. The verification must be performed to check if the attenuation at the inner peak of $\overline{uu}$ is not higher than at the outer peak. Again, according to Dróżdż \emph{et al.} \cite{Drozdz2024meas}, the attenuation in $\Delta\overline{uu}^+$ at the outer peak is twice as high at the inner peak, but still, as the $l^+$ is relatively short, the attenuation is within the uncertainty range, which in the aspect of the wavy wall termination, it proves that the location is correct.

The attenuation is relatively high for the two remaining locations $x = 0.3$ and $0.5$ m, as $l^+$ is 38 and 34, respectively. Assuming linear attenuation with $l^+$, it can be said that both are at the attenuation level of 40\% of those presented in Ref. \cite{Drozdz2024meas}. Then, the outer peak of $\overline{uu}$ is attenuated at about 4\% (still in uncertainty range) and as high as 16\% in the inner peak of $\overline{uu}$.

\bibliographystyle{aipnum4-1}
\bibliography{biblio}% Produces the bibliography via BibTeX.

\end{document}